\documentclass[prl,twocolumn,showpacs,amsmath,amssymb,superscriptaddress]{revtex4-1}
\usepackage{graphicx}
\usepackage{hyperref}
\usepackage{breqn}
\usepackage{xcolor}
\usepackage{color}
\usepackage{booktabs}
\usepackage{float}
\usepackage{longtable}
\usepackage{epsfig}

\makeatletter
\let\cat@comma@active\@empty
\makeatother
\begin{document}

\title{Efficient Electrical Spin-Splitter Based on  Non-Relativistic Collinear Antiferromagnetism}

\author{Rafael Gonz\'{a}lez-Hern\'{a}ndez }
\affiliation{Grup de Investigaci\'{o}n en F\'{i}sica Aplicada, Departamento de F\'{i}sica, Universidad del Norte, Barranquilla, Colombia}
\affiliation{Institut f\"ur Physik, Johannes Gutenberg Universit\"at Mainz, D-55099 Mainz, Germany}
\author{Libor \v{S}mejkal}
\affiliation{Institut f\"ur Physik, Johannes Gutenberg Universit\"at Mainz, D-55099 Mainz, Germany}
\affiliation{Institute of Physics, Czech Academy of Sciences, Cukrovarnick\'{a} 10, 162 00 Praha 6 Czech Republic}
\affiliation{Faculty of Mathematics and Physics, Charles University in Prague, Ke Karlovu 3, 121 16 Prague 2, Czech Republic}
\author{Karel V\'yborn\'y}
\affiliation{Institute of Physics, Czech Academy of Sciences, Cukrovarnick\'{a} 10, 162 00 Praha 6 Czech Republic}
\author{Yuta Yahagi}
\affiliation{Department of Applied Physics, Tohoku University, Sendai, Japan}
\author{Jairo Sinova}
\affiliation{Institut f\"ur Physik, Johannes Gutenberg Universit\"at Mainz, D-55099 Mainz, Germany}
\affiliation{Institute of Physics, Czech Academy of Sciences, Cukrovarnick\'{a} 10, 162 00 Praha 6 Czech Republic}
\author{Tom\'a\v{s}  Jungwirth}
\affiliation{Institute of Physics, Czech Academy of Sciences, Cukrovarnick\'{a} 10, 162 00 Praha 6 Czech Republic}
\affiliation{School of Physics and Astronomy, University of Nottingham, Nottingham NG7 2RD, United Kingdom}
\author{Jakub  \v{Z}elezn{\'y}}
\affiliation{Institute of Physics, Czech Academy of Sciences, Cukrovarnick\'{a} 10, 162 00 Praha 6 Czech Republic}

\date{\today}

\begin{abstract}
Electrical spin-current generation is among the core phenomena driving the field of spintronics. Using {\em ab initio} calculations we show that a room-temperature metallic collinear antiferromagnet RuO$_2$ allows for highly efficient spin-current generation, arising from anisotropically-split bands with conserved up and down spins along the N\'eel vector axis. The zero net moment antiferromagnet acts as an electrical spin-splitter with a 34$^\circ$ propagation angle between spin-up and spin-down currents. Correspondingly, the spin-conductivity is a factor of three larger than the record value from a survey of 20,000 non-magnetic spin-Hall materials. We propose a versatile spin-splitter-torque concept utilizing antiferromagnetic RuO$_2$ films interfaced with a ferromagnet.

\end{abstract}

\maketitle

Longitudinal spin-polarized current applied in an out-of-plane direction from a reference to a recording ferromagnetic layer  is the basis of commercial, spin-transfer-torque (STT) magnetic random access memories (MRAMs) \cite{Ralph2008,Brataas2012}.  Here the spin-current is due to ferromagnetic exchange-splitting, i.e., is odd under time-reversal ($\cal{T}$). An alternative example is  a transverse spin-current injected by an applied in-plane charge current from a non-magnetic relativistic spin-Hall polarizer \cite{Sinova2015} to the recording magnet in spin-orbit-torque (SOT) devices \cite{Manchon2019}.  Since this $\cal{T}$-even spin-current does not require a magnetic order, the family of potentially suitable materials is significantly broader. However, this is at the expense of relying on the relativistic spin-orbit coupling, which is typically much weaker than the exchange-coupling. Moreover, the same spin non-conserving nature of the spin-orbit coupling that allows for the spin-current generation diminishes the spin diffusion length and by this further limits theoretical understanding and practical utility of the relativistic charge-spin conversion phenomena \cite{Manchon2019}. 

The family of materials suitable for spin-current generation has been recently expanded by including non-collinear antiferromagnets \cite{Zelezny2017a,Zhang2017b,Kimata2019a,Mook2020}. Remarkably, despite the vanishing net magnetic moment, they allow for  the $\cal{T}$-odd spin-currents in analogy to ferromagnets.  However, the non-collinear magnetic order also generates spin-orbit splitting and spin texture in the electronic bands which resemble the relativistic spin-orbit coupling in spin-Hall materials. As a result, spin conservation is broken in non-collinear antiferromagnets even in the non-relativistic limit \cite{Zelezny2017a,Zhang2017b}. 

Here we show based on {\em ab initio} calculations in RuO$_2$ that large $\cal{T}$-odd spin-currents can be induced  by collinear antiferromagnetism without relativistic spin-orbit coupling. This mechanism of spin-current generation in a zero net-moment system accompanied by no spin-loss is a direct consequence of the anisotropic band spin-splitting, schematically illustrated in Fig.~1(a-c). It arises from the collinear exchange coupling between magnetic (Ru) atoms placed in locally anisotropic crystalline environments of the non-magnetic (O) atoms \cite{Smejkal2020,Ahn2019}. While the mechanism is applicable in a range of inorganic and organic collinear antiferromagnets 
\cite{Naka2019,Hayami2019,Yuan2020}, here we focus on RuO$_2$ which is a prominent, metallic room-temperature member \cite{Berlijn2017a,Zhu2018,Smejkal2020,Ahn2019,Hayami2019,Feng2020a} of a textbook family of rutile antiferromagnets \cite{Tinkham2003}. 

 \begin{figure}[h]
	\centering
	\includegraphics[width=0.490 \textwidth]{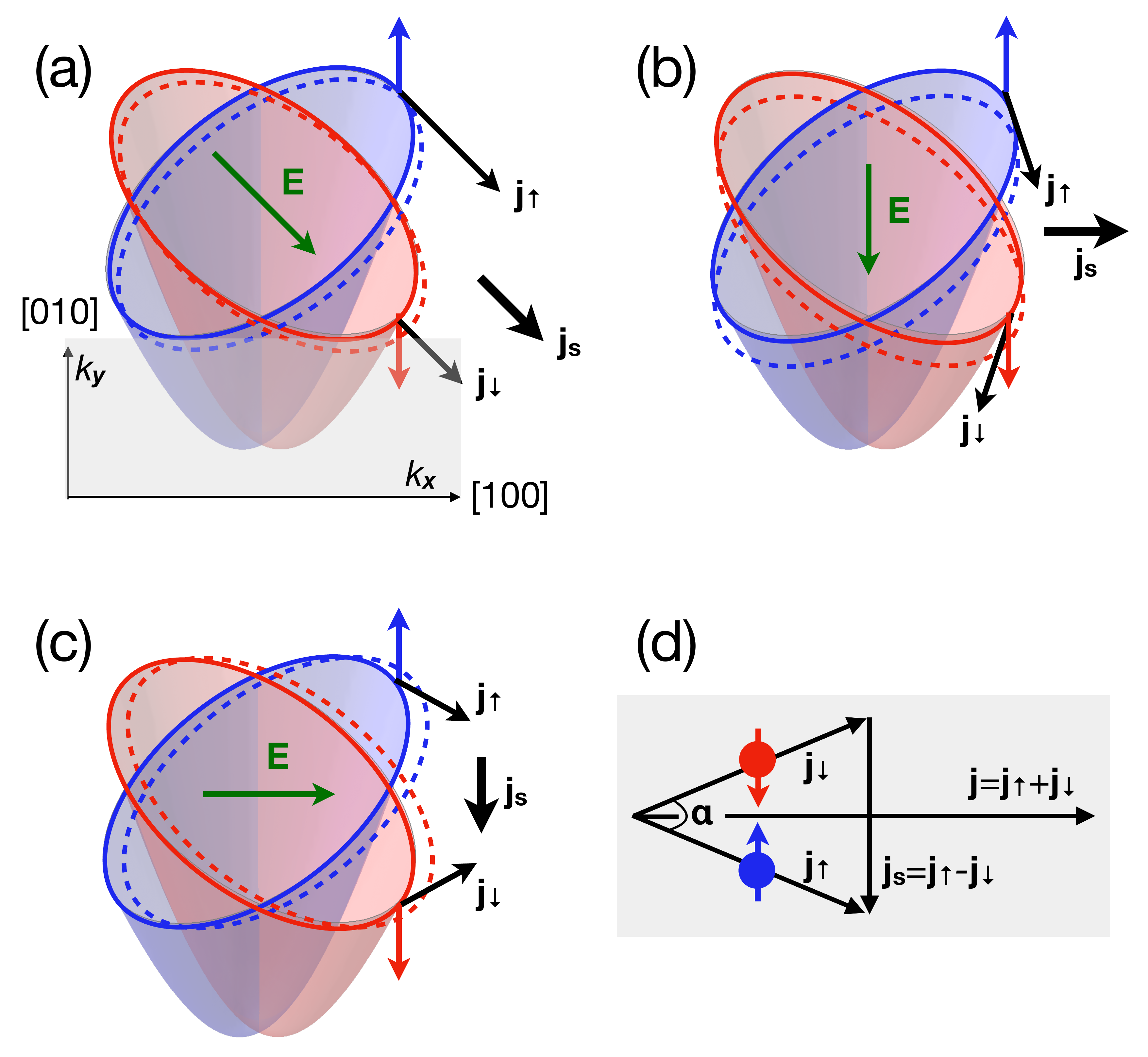} 
	\caption{Schematics of the anisotropic band splitting due to collinear antiferromagnetism and of the spin-splitter effect. (a) For an electric field {\bf E} applied along one of the $\langle 110\rangle$ axes, the spin-up and spin-down charge currents are parallel but of different magnitudes due to the band anisotropies. As a result, the longitudinal charge current is spin-polarized. (b,c) For {\bf E} along one of the $\langle 100\rangle$ axes, the spin-up and spin-down charge currents combine in an unpolarized longitudinal charge current and in a pure transverse spin-current. (d) In the absence of relativistic spin-orbit coupling, spin is conserved and the spin-up and spin-down charge currents are split by an angle  which is determined by the bands' anisotropy. In RuO$_2$, the angle $\alpha\approx34^\circ$.} 
	\label{fig:sc_origin}
\end{figure}

Recent theoretical and experimental works have demonstrated that the collinear antiferromagnetism accompanied by the unconventional anisotropic spin-splitting is also at the origin of the observed charge (spontaneous) Hall effect in RuO$_2$, although for this transport phenomenon the relativistic spin-orbit coupling is still essential \cite{Smejkal2020,Feng2020a}. The spin-current generation effect explored in this Letter is derived directly and solely from the anisotropic spin-splitting of the non-relativistic electronic structure.  We confirm that the calculated large conversion ratio between the charge current and the transverse pure spin-current is only weakly affected by the spin-orbit coupling present in the material. As shown schematically in Fig.~1(d), the collinear antiferromagnet acts as a direct splitter channeling the conserved up and down-spin electrons at a mutual angle of 34$^\circ$.  Based on our calculations presented below and referring to the earlier synthesized thin-film materials \cite{Feng2020a}, we  propose  a concept of a spin-splitter-torque (SST) in which an applied in-plane electrical current injects the spin-conserving pure spin-current polarized along the N\'eel vector in the out-of-plane direction from the RuO$_2$ film into a recording ferromagnetic layer. We conclude the Letter by illustrative examples of collinear antiferromagnets in which the non-relativistic spin-splitter effect is absent, and discuss our results in the context of physical phenomena recently associated with a term magnetic spin Hall effect \cite{Zelezny2017a,Kimata2019a,Mook2020}.

In our calculations, we use the density-functional theory (DFT) framework as implemented in the Vienna \textit{ab-initio} simulation package (VASP) \cite{Kresse1996a} within the  GGA+U approximation \cite{Berlijn2017a}. We perform both non-relativistic calculations and calculations with the relativistic spin-orbit coupling self-consistently included.  Electron wave functions are expanded in plane-waves up to a cut-off energy of 500~eV and a grid of 12x12x16 $k$-point is used to sample the irreducible Brillouin zone. 

\begin{figure*}[ht]
	\centering
	\includegraphics[width=1.0\textwidth]{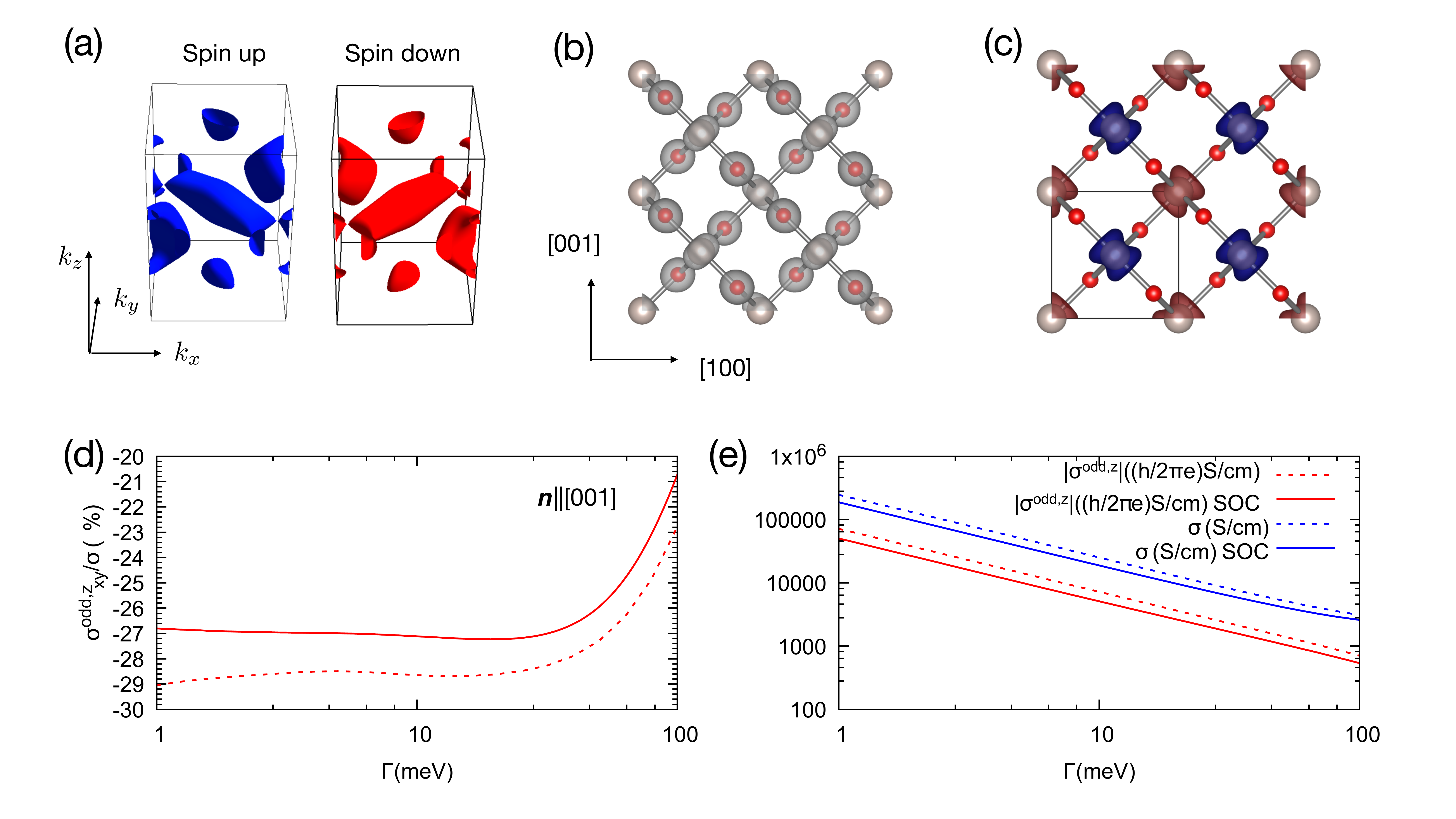}
	\caption{Electrical spin-splitter effect in RuO$_{2}$  from {\em ab initio} calculations. (a) Spin up and down Fermi surfaces. (b) Anisotropic electronic densities. (c) Anisotropic magnetization densities calculated without spin-orbit coupling \cite{Smejkal2020}. (d) $\cal{T}$-odd charge-spin conversion ratio calculated with and without (solid and dashed line) spin-orbit coupling and the N\'eel vector along [001] direction.  (e) Comparison of the corresponding longitudinal charge conductivity and transverse odd spin conductivity calculated with and without (solid and dashed lines) spin-orbit coupling.}
	\label{Fig2}
\end{figure*}%

In Fig.~2(a-c) we plot the DFT Fermi surfaces and charge and magnetization densities in RuO$_2$. We observe the correspondence between the momentum space and real space representations of the spin-splitter concept. The real space counterpart of the anisotropic band spin-splitting are the two intertwined sublattices with their respective anisotropies rotated by 90$^\circ$ (Fig.~2(b)) and one carrying primarily spin-up while the other one spin-down electrons (Fig.~2(c)). The spin-splitter angle is governed by the strength of these sublattice crystalline anisotropies which are high in RuO$_2$ and which give additional parameters like spin-orbit coupling or disorder a minor role. To quantify this, we evaluate  the charge to spin-current conversion ratio  within the linear response theory using the Kubo formula in the approximation of the constant scattering-rate $\Gamma$, as implemented in the Wannier-Linear-Response code \cite{Zelezny2017a,Jakubcode}. 

The spin-conductivity is described by a  tensor $\sigma^a_{bc}$, where $a$ corresponds to the spin-polarization of the spin-current, $b$ to the direction of spin-current flow and $c$ to the direction of the applied electric field. The Kubo formula within the constant $\Gamma$ approximation can be split into the  $\cal{T}$-odd contribution, 
 \begin{eqnarray}
	& &\sigma^{{\rm odd},a}_{bc} =  \\
	&-&\frac{e\hbar}{V\pi} \text{Re}\sum_{\textbf{k}, m, n}
	\frac{\left\langle u_{n}(\textbf{k})\vert \hat{J}^a_b \vert u_{m}(\textbf{k}) \langle u_{m}(\textbf{k})\vert \hat{v}_c \vert u_{n}(\textbf{k})  \right\rangle \Gamma^{2}}{(\left(E_{F}-E_{n}(\textbf{k}))^{2}+\Gamma^{2})(E_{F}-E_{m}(\textbf{k})\right)^{2}+\Gamma^{2})}\,, \nonumber
	\label{boltzmann}
\end{eqnarray}
and the $\cal{T}$-even contribution given in the $\Gamma\rightarrow0$ limit by,
\begin{eqnarray}
   & &\sigma^{{\rm even},a}_{bc}=  \\
   & -&\frac{2e{\hbar}}{V} \text{Im}\sum_{\textbf{k},m\neq n}^{\substack{ \\ n\ \text{occ.}\\ m\ \text{unocc.}}}
	\frac{\left\langle u_{n}(\textbf{k})\vert \hat{J}^a_b \vert u_{m}(\textbf{k}) \langle u_{m}(\textbf{k})\vert \hat{v}_c \vert u_{n}(\textbf{k})  \right\rangle}{\left(E_{n}(\textbf{k})-E_{m}(\textbf{k})\right)^{2}} \,. \nonumber
	\label{intrinsic}
\end{eqnarray}
Here  $u_n(\textbf{k})$ are the Bloch functions of a band $n$, \textbf{k} is the Bloch wave vector, $\varepsilon_{n}$(\textbf{k}) is the band energy, $E_F$ is the Fermi energy, $\hat{v}$ is the velocity operator,   and the spin-current operator $\hat{J}^a_b = \frac{1}{2}\{\hat{s}_a,\hat{v}_b\}$. In order to evaluate the Kubo formula, we construct  an effective tight-binding Hamiltonian in the maximally localized Wannier basis \cite{Mostofi2008} as a post-processing step of the DFT calculations. For the integration, we use a dense $320^3$ \textbf{k}-mesh.

For small $\Gamma$, the Kubo formula (1) for $\sigma^{{\rm odd},a}_{bc}$, which describes our spin-splitter effect, scales as $1/\Gamma$ and is equivalent to the semiclassical Boltzmann equation with constant scattering rate. In RuO$_2$, $\Gamma=6.6$~meV corresponds to an average experimental value of the room-temperature charge conductivity $\sigma=2.8\times 10^4$~$\Omega^{-1}$cm$^{-1}$ \cite{Ryden1970}. This value of $\Gamma$ is safely below the upper limit ($\sim 100$~meV)  of the $1/\Gamma$ scaling of $\sigma^{{\rm odd},a}_{bc}$ and $\sigma$ ($=\sigma_{xx}$), as shown in Fig.~2(e). The $\cal{T}$-odd charge-spin conversion ratio  $\sigma^{{\rm odd},a}_{bc}/\sigma$ is then only weakly dependent on disorder scattering, as anticipated above and confirmed numerically in Fig.~2(d). 

While  $\sigma^{{\rm odd},a}_{bc}$ diverges for $\Gamma\rightarrow0$,  $\sigma^{{\rm even},a}_{bc}$ is finite in the clean limit and described by the intrinsic  Kubo formula (2). It implies that the corresponding conversion ratio $\sigma^{{\rm even},a}_{bc}/\sigma\rightarrow0$ for $\Gamma\rightarrow0$. Since the intrinsic contribution often dominates in the favorable, strongly spin-orbit coupled materials \cite{Sinova2015}, this is a significant limiting factor of the $\cal{T}$-even charge-spin conversion in metals. The $\cal{T}$-odd part circumvents this limitation in metallic systems and it also circumvents the spin-loss inherent to the $\cal{T}$-even spin conductivity \cite{Zelezny2017a,Zhang2017b}. This is highlighted in Fig.~2(d) where we show that the  effect of the relativistic spin-orbit coupling on the $\cal{T}$-odd charge-spin conversion ratio is weak, despite the presence of significant spin-orbit coupling in the material due to the heavy element Ru (cf. Fig.~2(e)). Fig.~2(d) thus confirms the efficiency and the non-relativistic spin-splitter nature of the phenomenon in RuO$_2$. The calculated charge-spin conversion ratio $\sigma^{{\rm odd},a}_{bc}/\sigma\approx28\%$ which corresponds to the angle between the spin-up and spin-down transport channels of 34$^\circ$ (Fig.~1(d)). We point out that our spin-splitter conversion ratio is significantly larger than the experimentally reported charge-spin conversion ratios (angles) in the most extensively explored spin-Hall metal Pt, and comparable to the record metallic spin-Hall conversion material $\beta$-W \cite{Sinova2015}.

The $\sigma^{{\rm odd},a}_{bc}$ calculations shown in Figs.~2(d,e) are for the $b=x$ ($[100]$) and $c=y$ ($[010]$) crystal directions in tetragonal RuO$_2$. Without relativistic spin-orbit coupling, $\sigma^{{\rm odd},a}_{xy}=\sigma^{{\rm odd},a}_{yx}$ (cf. Figs.~1(b,c)) are the only non-zero $\cal{T}$-odd spin-conductivity components in the material, which can be formally verified using the symmetry group analysis \cite{Zhang2017b,symcode}. The spin-polarization of the non-relativistic spin-current is along the spin quantization axis of the electronic structure, i.e., is aligned with the axis of the N\'eel vector $n$.  Calculations in  Figs.~2(d,e) were done for $n\parallel [001]$ and we, therefore, plot the  $a=z$ ($[001]$) component of $\sigma^{{\rm odd},a}_{xy}$. Note that when spin-orbit coupling  is included, more components of $\sigma^{{\rm odd},a}_{bc}$ are allowed by symmetry, as summarized in the Supplementary information.

\begin{figure}[ht]
	\centering
	\includegraphics[width=0.490\textwidth]{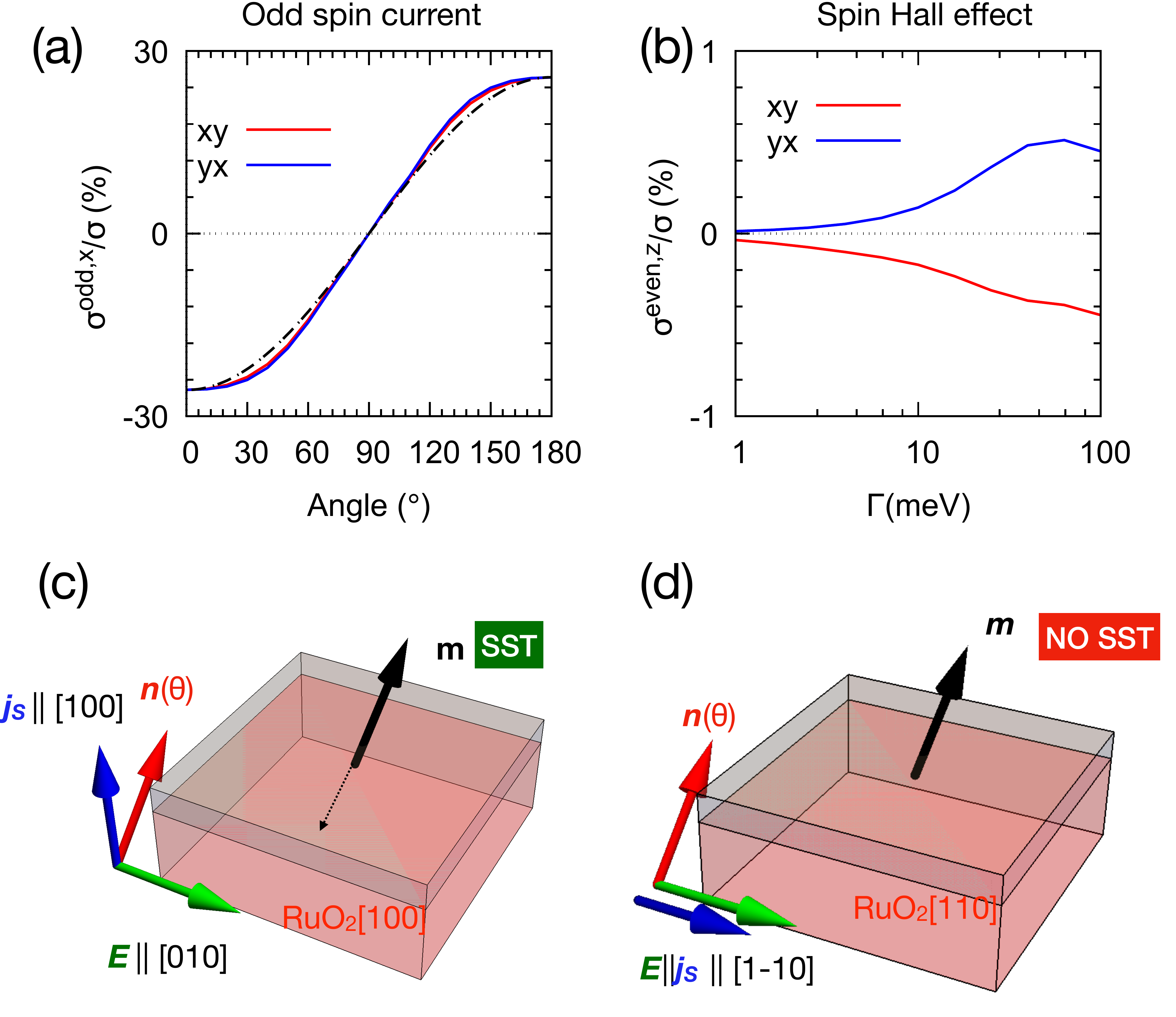}
	\caption{Spin-splitter-torque concept. (a) $\cal{T}$-odd spin-splitter charge-spin conversion ratio as a function of the in-plane N\'eel vector angle highlighting that the polarization of the spin-current follows the N\'eel vector. (b) $\cal{T}$-even spin-Hall charge-spin conversion ratio as a function of the scattering rate. (c,d) Schematics of antiferromagnetic (red) / ferromagnetic (gray) bilayer with two different orientations of the RuO$_{2}$ crystal. Configuration (c) gives a spin-current in the antiferromagnet propagating in the out-of-plane direction which can generate a spin-splitter-torque on the magnetization of the ferromagnet. Configuration (d) gives an in-plane spin-current and, therefore, generates no spin-splitter-torque.}
	\label{Fig3}
\end{figure}%

In Fig.~3(a) we show that the spin-orbit coupling has a weak effect on the charge-spin conversion ratio also when the N\'eel vector is in the $x-y$ plane of RuO$_2$. The cosine dependence of the plotted $\sigma^{{\rm odd},x}_{xy}$ as a function of the in-plane N\'eel vector angle measured form the $x$-axis again reflects that the spin polarization of the leading non-relativistic contribution to the spin-current is aligned with $n$.  In Fig.~3(b) we show for comparison the $\cal{T}$-even charge-spin conversion ratio for the spin-Hall  component $\sigma^{{\rm even},z}_{xy}=-\sigma^{{\rm even},z}_{yx}$ in RuO$_2$. For the experimental value of $\Gamma=6.6$~meV, the non-relativistic $\cal{T}$-odd  spin-splitter  outperforms  by two orders of magnitude the $\cal{T}$-even spin-Hall effect, which is of relativistic origin in RuO$_2$  as in non-magnets \cite{Sinova2015} (see Supplementary information). We also point out here that our calculated non-relativistic $\cal{T}$-odd spin-conductivity in RuO$_2$ is a factor of three larger than the record value of the relativistic $\cal{T}$-even spin-Hall conductivity from a survey of 20,000 non-magnetic materials \cite{Zhanga}.

We now discuss the utility of the spin-splitter concept in spin-torque structures. We recall that STT based on the $\cal{T}$-odd spin-polarized current mechanism employed in commercial MRAMs suffers from some limitations given by the two-point out-of-plane geometry of both the electrical STT writing and the electrical readout by tunneling magnetoresistance. One particular challenge in STT-MRAMs is to balance a sufficiently large gap between writing and readout currents with strong enough readout signals and with writing currents kept safely below the breakdown threshold of the tunnel barrier \cite{Ralph2008,Brataas2012}.  Research and development of SOT MRAMs based on the $\cal{T}$-even spin-Hall currents is fuelled in large part by the possibility to employ a three-point geometry with an in-plane ohmic path for the writing electrical current. However, they rely on the spin-orbit coupling which is spin non-conserving and more subtle than the ferromagnetic exchange. Both these features can limit the SOT efficiency.
Our SST, illustrated in Fig.~3(c), shares the versatility of STT offered by the control of the $\cal{T}$-odd spin current by the magnetic order. Simultaneously, it does not inherit the  problems of STT associated with the applied out-of-plane electrical writing current. Instead, SST shares the in-plane electrical writing geometry of SOT while circumventing the spin-loss limitations of the $\cal{T}$-even spin-currents. 

In Fig.~3(c) we consider a RuO$_2$/ferromagnet bilayer grown along the [100] crystal axis, as reported in Ref.~\cite{Feng2020a}. This is the optimal geometry for SST. Here the electric field applied along the in-plane [010] axis generates the pure non-relativistic spin-current along the out-of-plane [100] direction with the polarization of the spin-current determined by the N\'eel vector (cf. Fig.~1(b)). In stoichiometric  RuO$_2$, the N\'eel vector easy-axis is along the [001]  direction \cite{Berlijn2017a,Smejkal2020,Feng2020a} which implies efficient  SST for switching of an in-plane magnetized ferromagnet interfaced with RuO$_2$. 

The easy axis in RuO$_2$ can be rotated into the (001)-plane by off-stoichiometry or alloying (with e.g. Ir) \cite{Smejkal2020}  in which case SST is optimized for switching an out-of-plane magnetized ferromagnet. On the other hand, a RuO$_2$ film grown along the [110] axis \cite{Feng2020a} only allows for generating a longitudinal spin-polarized current propagating along the applied in-plane ($\parallel [1\bar{1}0]$) electric field. This is seen in the model anisotropic spin-split bands in Fig.~1(a) and can be formally obtained from the calculated  $\sigma^{{\rm odd},a}_{xy}=\sigma^{{\rm odd},a}_{yx}$ components by rotating the cartesian coordinates by 45$^\circ$. The geometry precludes the SST switching of the interfaced ferromagnet (Fig.~3(d)), a feature that can be used to test the SST phenomenology in experiment.  Note that a longitudinal spin-polarized current propagating along an applied out-of-plane ($\parallel [110]$) electric field could be used for the STT switching of the adjacent ferromagnet.  We also recall here that the $\cal{T}$-odd spin-current  is predicted  to be two-orders of magnitude larger than the $\cal{T}$-even  spin current (cf. Figs.~3(a,b)). Experimentally disentangling the SST in RuO$_2$/ferromagnet bilayers grown along the [100] crystal axis from the SOT contribution should, therefore, be readily feasible. 

We conclude our discussion with a few remarks on the spin-splitter effect in broader materials and charge-spin conversion contexts. The spin-splitter effect in which an electrical current generates a pure conserving spin-current requires a collinear antiferromagnet. However, not all collinear antiferromagnets have the effect allowed by symmetry. The spin-conductivity  is invariant under spatial translation ($t$) or inversion ($\cal{P}$), and  the $\cal{T}$-odd spin-conductivity requires $\cal{T}$-symmetry breaking. This implies that it is excluded in systems with combined $\it{t}\cal{T}$ or $\cal{PT}$ symmetries. FeRh or CuMnAs are examples of materials used in spintronics research of metallic collinear antiferromagnets \cite{Jungwirth2016} falling into this high-symmetry category, and which therefore do not show this effect. 

 \begin{figure}[h]
	\centering
	\includegraphics[width=0.490 \textwidth]{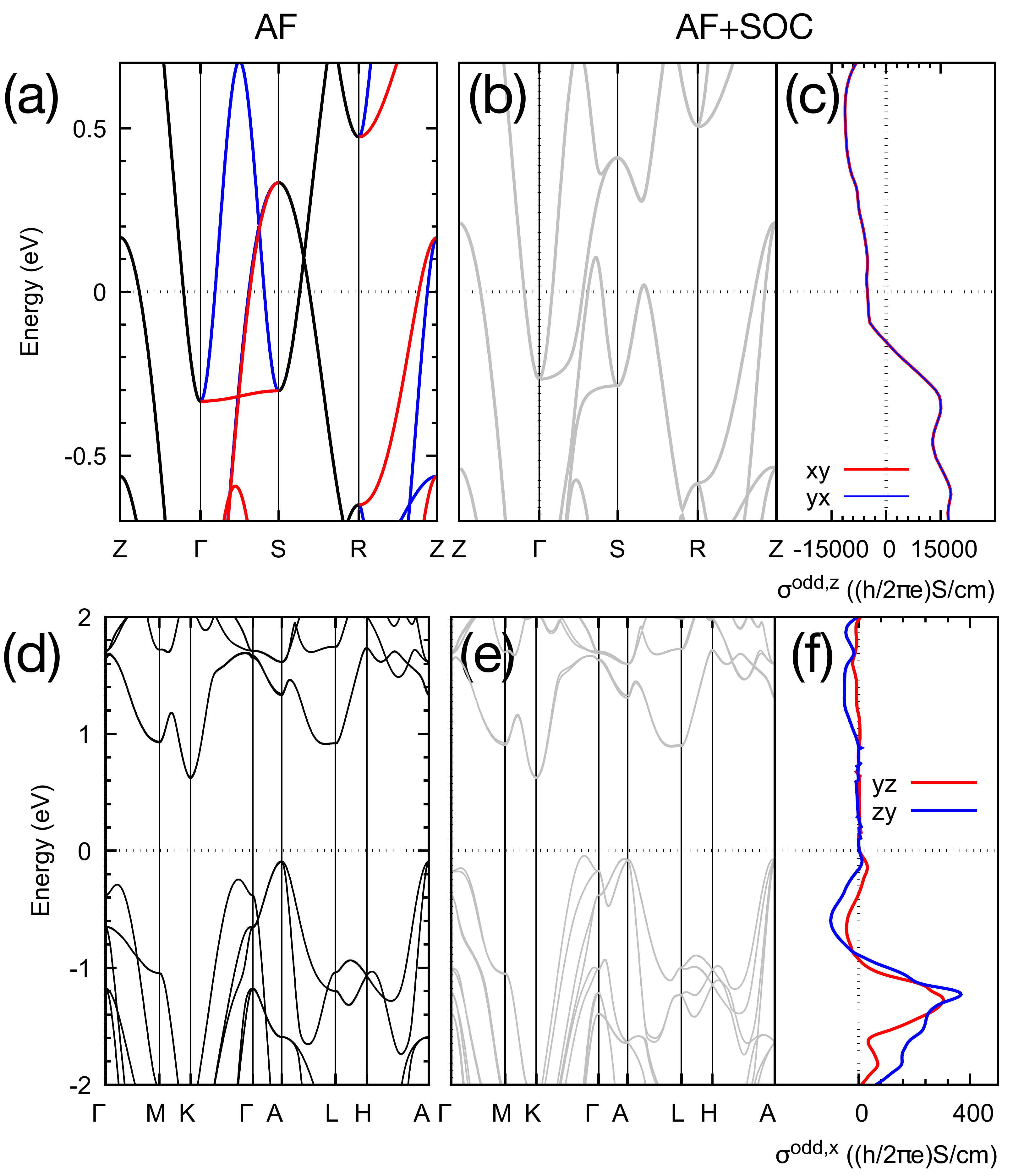} 
	\caption{Energy bands and $\cal{T}$-odd spin-conductivity  in collinear antiferromagnets calculated from first-principles. (a) Electronic bands of  RuO$_2$ calculated without spin-orbit coupling  for the N\'eel vector oriented along the [001] easy axis. Red and blue lines highlight the splitting of the spin-up and spin-down bands while black lines correspond to spin degenerate bands. (b) Same as (a) with spin-orbit coupling. (c) Energy dependent $\cal{T}$-odd spin-conductivity calculated with spin-orbit coupling and $\Gamma$=10 meV.  (d,e) Same as (a,b) for MnTe, Hubbard $U=3.03$~eV, and the N\'eel vector along the [1$\overline{1}$00] axis. (f) Same as (c) for MnTe and  $\Gamma$=45 meV.} 
	\label{fig4}
\end{figure}

MnTe, another prominent material in antiferromagnetic spintronics research \cite{Kriegner2016}, is a different case. This collinear semiconducting antiferromagnet has both the $\it{t}\cal{T}$ and $\cal{PT}$ symmetries broken and the $\cal{T}$-odd spin-conductivity is allowed in this material, as in all crystals breaking $\it{t}\cal{T}$ and $\cal{PT}$ (see Supplementary information). However, it requires spin-orbit coupling while the non-relativistic spin-splitter effect is absent in MnTe. This can be again shown formally by the symmetry group analysis \cite{symcode} while in Fig.~4 we provide a more microscopic physical insight. In Fig.~4(a) we highlight the anisotropic splitting of the spin-up and spin-down states in the non-relativistic electronic structure responsible for the spin-splitter effect in RuO$_2$\cite{Smejkal2020}. (Note that much of the character of the spin-split bands is preserved in the presence of spin-orbit coupling, as shown in Fig.~4(b).)  MnTe  has a more symmetric arrangement of non-magnetic atoms in the lattice leading to spin-degenerate non-relativistic bands along high symmetry lines, as shown in Fig.~4(d)\cite{Yin2019}. 
The $\cal{T}$-odd spin-conductivity occurs only when the symmetry of the system is lowered by including spin-orbit coupling. We note that the relativistic $\cal{T}$-odd spin-conductivity in MnTe is again one--two orders of magnitude weaker than the non-relativistic counterpart in  RuO$_2$ (cf. Figs.~4(c) and 4(f)). 

Finally, we point out that both $\cal{T}$-odd and $\cal{T}$-even spin conductivity tensors can have symmetric and antisymmetric components (i.e. also longitudinal and transverse components) for a general crystal symmetry. This is in contrast to the charge conductivity tensor whose $\cal{T}$-even  part is purely symmetric and $\cal{T}$-odd part purely antisymmetric. As a result, there is some ambiguity in the terminology of the charge-spin conversion phenomena \cite{Zelezny2017a,Zhang2017b,Kimata2019a,Mook2020,Wimmer2015,Seemann2015}. Traditionally, when focusing on isotropic models, the term spin Hall effect has been more narrowly reserved to the antisymmetric part of the $\cal{T}$-even spin conductivity \cite{Sinova2015}. However, in lower symmetry crystals, the symmetric $\cal{T}$-even spin conductivity can be  also allowed. The term spin Hall effect has been then used in a broader sense to capture the full $\cal{T}$-even spin conductivity tensor, including its antisymmetric and symmetric components \cite{Zhang2017b,Mook2020}. Similar ambiguity exists with the recently introduced term magnetic spin Hall effect in the context of studies of the $\cal{T}$-odd spin conductivity in non-collinear antiferromagnets \cite{Zelezny2017a,Kimata2019a,Mook2020}. Our non-relativistic spin-splitter effect in RuO$_2$ is described by the symmetric $\cal{T}$-odd spin conductivity and can, therefore, be considered to fall within the family of magnetic spin Hall effects in the broader sense of the term.

\section*{Acknowledgments}
We acknowledge funding form the Czech Science Foundation Grant No. 19-18623Y,  the Ministry of Education of the
Czech Republic Grants No. LM2018096, LM2018110, and LNSM-LNSpin, and  the EU FET Open RIA Grant No. 766566. Funded by the Deutsche Forschungsgemeinschaft (DFG, German Research Foundation) - TRR 173 – 268565370 (project A03). J. \v{Z}. acknowledges support from the Institute of Physics of the Czech Academy of Sciences and the Max Planck Society through the Max Planck Partner Group programme. This work was supported by The Ministry of Education, Youth and Sports from the Large Infrastructures for Research, Experimental Development and Innovations project „e-Infrastructure CZ – LM2018140“. The authors gratefully acknowledge the computing time granted on the supercomputer Mogon at Johannes Gutenberg University Mainz (hpc.uni-mainz.de) and the support of Alexander Von Humboldt Foundation.

\bibliography{Refs,other,MSHE}

\end{document}